\documentclass[aps,pra,onecolumn,superscriptaddress,11pt]{revtex4}

\usepackage{latexsym}
\usepackage{amsfonts}
\usepackage{graphicx}
\usepackage{dcolumn}
\usepackage{url}

\newcommand{\rme}{{\mathrm{e}}}

\newcommand{\rmd}{{\mathrm{d}}}
\newcommand{\rmi}{{\mathrm{i}}}

\newcommand{\RR}{{\mathbb R}}

\newcommand{\ZZ}{{\mathbb Z}}

\newcommand{\cA}{{\mathcal A}}

\newcommand{\cN}{{\mathcal N}}

\newcommand\abs[1]{{| #1 |}}

\newcommand\norm[1]{{\| #1 \|}}

\begin{document}
\def\Marseille{UMR 7345 CNRS, Aix-Marseille Universit\'e, 
campus Saint-J\'er\^ome, case 321, \\ 
av.\ esc.\  Normandie-Niemen, FR-13397 Marseille cedex 20}
\title{Vlasov equation and $N$-body dynamics  \\
     \small{How central is particle dynamics to our understanding of plasmas~?}
    }
\author{Yves Elskens, D~F Escande and F Doveil}
\affiliation{\Marseille}
\email{yves.elskens@univ-amu.fr, dominique.escande@univ-amu.fr, fabrice.doveil@univ-amu.fr}

%
\begin{abstract}
Difficulties in founding microscopically the Vlasov equation  
for Coulomb-interacting particles are recalled for both 
the statistical approach (BBGKY hierarchy and Liouville equation on phase space) 
and the dynamical approach (single empirical measure 
on one-particle $(\mathbf{r},\mathbf{v})$-space).
The role of particle trajectories (characteristics) in the analysis of 
the partial differential Vlasov--Poisson system is stressed.
Starting from many-body dynamics, 
a direct derivation of both Debye shielding and collective behaviour is sketched.
      {05.20.Dd}  {Kinetic theory}  \newline 
      {45.50.-j}  {Dynamics and kinematics of a particle and a system of particles}   \newline 
      {52.35.Fp}  {Plasma: electrostatic waves and oscillations}
\end{abstract} 
\maketitle

\section{Introduction}
\label{sec:Intro}

Plasmas are many-body systems, in which the long-range Coulomb interactions play the leading role. 
Because of the long range coupling, each particle interacts with many other particles, 
so that it seems natural to attempt describing this medium in a mean-field limit. 
This approach has proved successful as it provides a theoretical frame for computing various phenomena, 
in quantitative agreement with physical observation -- for natural and for laboratory plasmas. 

However, because the Coulomb field diverges strongly near its source, 
a complete derivation of the Vlasov--Poisson equations from first principles in Coulomb systems 
is yet to be constructed (see \cite{Kie08,Kie14} and references therein). 
We recently tried to recover some vlasovian results starting from the $N$-body picture, 
and taking the $N \to \infty$ limit after solving approximately the dynamics \cite{EDE13}. 
While this yields an alternative derivation for some Vlasov-based results, 
it does not provide a direct validation of the Vlasov--Poisson system in the continuum limit
-- actually, it bypasses the Vlasov picture by leading also easily to the analysis of collisions, and 
shows moreover how Debye shielding alters the Poisson description of the interactions \cite{EED14}. 

In this paper, we discuss some of the respective merits 
of the partial differential equation viewpoint inherent to the Vlasov equation and 
of the ordinary differential equation viewpoint of many-body Coulomb force dynamics. 
We hope this discussion will help the reader to grasp how delicate foundational issues  
the Vlasov equation involves. 
While our selection of topics is bound to remain far from exhaustive, 
we hope it will refresh the reader's view of the Vlasov equation. 

Specifically, Sect.~\ref{sec:Status} focuses on the meaning of the entity obeying the Vlasov equation,
namely the distribution function. 
Section~\ref{sec:VP} comments on the role of particle trajectories 
in the theory of Vlasov--Poisson systems. 
Section~\ref{sec:Nbody} sketches a specific instance where attention to trajectories 
provides better insight into well-known vlasovian phenomena 
-- this is the only section where a specific microscopic model (actually, jellium) is considered.

\section{Status of the distribution function and its evolution}
\label{sec:Status}

Major merits of the Vlasov equation include 
\begin{enumerate}
\item{\label{merit-1} 
  a rather successful application to the modeling of hot plasmas,
  }
\item{\label{merit-2}
   enabling physicists to use powerful\footnote{The standard toolbox 
       of the plasma physicist 
       would be dramatically damaged by the removal of partial differential equations
       and the associated techniques -- even the most ``elementary'' ones like separation of variables.
       }
  techniques of partial differential equations to address physical phenomena,
  }
\item{\label{merit-3} 
   contributions to progress in functional analysis\footnote{This progress 
      makes the modern theory of partial differential equations
      incommensurately more powerful, benefiting to many other fields.
      }
  motivated by challenges it raises,
  }
\item{\label{merit-4}
   stressing an interesting stage 
   (in order to formulate a physical problem, one needs not merely to write its equations
   but first to choose the mathematical entities called to play) 
   for describing the physics. 
   }
\end{enumerate}
Issue \ref{merit-1} indicates how much needed is a comprehensive foundation for this fruitful model. 
We address issue \ref{merit-2} (and indirectly issue \ref{merit-3}) in Sect.~\ref{sec:VP}. 
Let us stress in the current section the conceptually crucial issue \ref{merit-4}~: 
the central quantity in the classical theory of the Vlasov equation is 
a non-negative continuous function $f$ on Boltzmann's $\mu$-space \cite{PenroseBk}, 
$f(\mathbf{r},\mathbf{v},t)$, which in the simple Coulomb interaction case 
obeys a closed evolution equation\footnote{The Poisson equation is not an evolution equation~: 
   it simply provides an explicit relation expressing the electric field in terms of the current, instantaneous $f$ 
   -- in this sense, the electric field does not describe additional degrees of freedom,
   but is a mere subsidiary entity slaved to the particles. 
   }.

\subsection{Pedestrian approach}
\label{sec:Status-pedestrian}

Classical, continuously differentiable distribution functions $f(\mathbf{r},\mathbf{v},t)$ 
are often interpreted in terms of some averages. 
Since the particles are completely described by their positions and velocities, 
$(\mathbf{r}_j(t),\mathbf{v}_j(t))$,
one first associates with these data a ``spiky'' distribution, 
involving Dirac distributions on $(\mathbf{r},\mathbf{v})$ space. 
This is too wild an object for simple calculations, and one would gladly smoothe it.
Moreover, as the field acting on particles varies in space, 
distinct particles are subject to different accelerations, 
and their motions may significantly separate with time.
Yet, while particles drift apart from each other, other particles may come around them 
in such a way that the overall distribution does not seem to change much~:
recall a cloud made of many tiny drops, 
or the large-scale hydrodynamics of air made of many molecules. 

Simultaneously, the field generated by the particles varies wildly on microscopic scales,
so that one expects particle motions to be hardly accessible to the theory on these scales. 
Therefore, one would also gladly limit the particle motion analysis to large scales, 
over which the Coulomb field might appear smoother. 

A \emph{first smoothing procedure} is simply to replace the Coulomb interaction 
(with point sources) by a regularized, or mollified \cite{Neunzert84}, interaction, 
where the source of the Coulomb field is a sphere with finite charge density 
centered on the point particle. 
The force on a particle is then also computed by summing the electric field over the sphere. 
This is done in particle-based numerical schemes, and if one keeps a fixed regularizing form function,
one can even obtain the Vlasov equation from the $N$-body model rigorously (see e.g.\ \cite{Neunzert84}).
However, the derivation of the Vlasov equation depends crucially 
on the size of the mollifying sphere.\footnote{Physically, 
   one expects the classical model to fail anyway at describing 
   Coulomb-interacting particles at very short distances. 
   Indeed, the ``classical radius of the electron'' (at which scale the point particle model should break down)
   is far smaller than its Compton length, so that quantum effects are expected to imply a different modeling. 
   It may thus be considered physically sensible to model the $N$-body plasma with a ``quantum-mollified'' 
   Coulomb interaction (even for large, finite $N$). 
   This will seem even more harmless since 
   the quantum description of two-body interaction \cite{Bohm51} leads to a scattering cross section identical 
   to the classical Rutherford cross section if particles bear charges with equal signs -- 
   moreover, the same expression applies to particles with opposite signs, 
   for which the classical picture allows for arbitrarily close approaches.
   Besides, physicists most often consider plasmas 
   where typical binary collisions involve closest approaches 
   much further away than the Compton length,
   and molecular dynamics simulations simply integrate the Coulomb force for short range 
   while smoothing the interaction for longer range \cite{CouedelPrivate}.
   Yet the mathematical issue is important, for it should provide insight into more complicated problems too.}

A \emph{second procedure} to formalize the ``cloud'' picture is to deem irrelevant 
some subtleties of particle motion. 
Rather, one pays attention to the evolution of the particles 
currently in a ``mesoscopic'' domain $\Delta U_\mathbf{r}$
(with size $| \Delta U_\mathbf{r} | = \Delta x \Delta y \Delta z$), 
with velocities in a similar range $\Delta U_{\mathbf{v}}$
(with $| \Delta U_\mathbf{v} | = \Delta v_x \Delta v_y \Delta v_z$).
The distribution function is then used to compute the ``coarse-grained'' distribution 
$| \Delta U_\mathbf{r} \times \Delta U_\mathbf{v} |^{-1} \int_{\Delta U_\mathbf{r} \times \Delta U_\mathbf{v}}
     f(\mathbf{r},\mathbf{v},t) \, \rmd^3 \mathbf{r} \, \rmd^3 \mathbf{v}$,
where the range and domain of interest are large enough 
to contain so many particles (say $Q \gg 1$)
that their number would fluctuate moderately with time
(say on the smaller scale $Q^{1/2}$ 
if these fluctuations follow a central-limit type of scaling, 
or $Q^{2/3}$ for a surface-vs-volume scaling in position space,
or $Q^{5/6}$ for a boundary-vs-volume scaling in $(\mathbf{r},\mathbf{v})$-space).

To extract a smooth function $f$ from this fluctuating picture, 
a \emph{third procedure} consists in introducing an ``ensemble'' of realizations of the plasma, 
see e.g.\  \cite{Akhiezer,Ichi73,HaWa04}.
The function $f$ in the Vlasov equation is then viewed as the average of 
individual spiky distributions. 
As the force acting on a particle is due to the field generated by all other particles, 
this force is expressed as an integral over the distribution of those source particles,
from which the target particle is excluded. 
This leads to computing a field $\mathbf{E}(\mathbf{r}_1,t)$ 
from an integral over the two-particle joint distribution 
$f^{(2)} (\mathbf{r}_1, \mathbf{v}_1, \mathbf{r}_2, \mathbf{v}_2, t)$, 
in the spirit of derivations of the Boltzmann equation in gas theory.\footnote{Until 2013,
   rigorous derivations of the Boltzmann equation were limited 
   to short times (on the order of the mean free path) 
   or small initial data (ultimately expanding in vacuum without collision).
   This is a topic of hard and active research,
   see \cite{GalStRayTex,Pulvirenti13} 
   and references therein.}
Because of the long range nature of the Coulomb interaction, 
one expects particles to be ``almost'' independent and $f^{(2)}$ to almost factorize
-- otherwise, one should solve an evolution equation for $f^{(2)}$ 
where the source involves $f^{(3)}$, etc... 
The resulting set of equations, viz.\ the BBGKY hierarchy, 
stands at the core of the statistical approach below.\footnote{Note that, in 
   the $N$-body case, giving the spiky distribution amounts to locating all particle,
   and the joint distribution for pairs is uniquely determined by these data. 
   In contrast, given a smooth distribution $f^{(1)}$ on $\RR^6$,
   there exist infinitely many joint distributions $f^{(2)}$ on $R^{12}$ 
   such that 
   $\int f^{(2)} (\mathbf{r}_1, \mathbf{v}_1, \mathbf{r}_2, \mathbf{v}_2, t) 
                     \, \rmd^3 \mathbf{r}_2 \rmd^3 \mathbf{v}_2
      = f^{(1)} (\mathbf{r}_1, \mathbf{v}_1, t)$ 
   along with $f^{(2)} (1,2) = f^{(2)} (2,1)$. 
   The factorization $f^{(2)} (1,2) = f^{(1)}(1) f^{(1)}(2)$ may hold at initial time but is generally not
   preserved by the time evolution. 
   This ``propagation of (initial) chaos'' (or ``Kac property'') is an important, non-trivial issue,
   for understanding the $N \to \infty$ limit \cite{MiMouWenn}.}

Let us stress again that the usual conceptual setting for these derivations
involves probabilistic averages over ensembles to generate smooth functions. 
Yet a physical plasma is a single realisation of the possible plasmas considered in an ensemble. 
The particles in it do not respond to the average field generated by the ensemble,
and each particle follows a single, regular enough, trajectory (which is not a diffusion process). 

An additional difficulty met with continuous distribution functions $f$ solving the Vlasov equation
is the absence of H-theorem. Indeed, the Vlasov equation preserves 
all functionals of the form $\int_{\RR^6} G(f(\mathbf{r}, \mathbf{v})) \rmd^3 \mathbf{r} \rmd^3 \mathbf{v}$,
and evolutions towards a kind of equilibrium can only lead to the formation of 
finer and finer filaments rippling the surface representing $f$ 
over $(\mathbf{r}, \mathbf{v})$ space.\footnote{This  is also viewed 
   from the fact that the Vlasov equation for hamiltonian dynamics 
   ``shuffles'' (without altering their volumes) 
   the respective domains over which $f$ takes its various values, 
   i.e.\ at any time $t$ the set $\{ (\mathbf{r}, \mathbf{v}) : a < f(\mathbf{r}, \mathbf{v}, t) \leq b \}$ 
   has the same 6-dimensional volume (see \cite{HaWa04} for this shuffling image). 
   For the one-dimensional models, with $(\mathbf{r}, \mathbf{v}) = (x,v)$,
   volume preservation reduces to area preservation,
   which makes hamiltonian dynamics so different from dissipative ones. 
   Recall also how mixing occurs in non-diffusive, viscosity-free incompressible fluid flows.}
When such ripples become finer than a typical interparticle distance in the $N$-body system,
they lose physical significance \cite{FirpoPoleni}. 
However, the BBGKY evolution equations 
do not incorporate such a destruction of unphysical ripples.

In numerical simulations, this filamentation is a delicate issue.
On the one hand, modeling accurately the partial differential equation 
requires increasing computational power as filamentation proceeds.
On the other hand, the numerical smoothing due to various interpolation schemes is not granted 
to reproduce the physical smoothing of the distribution function due to 
phenomena not included in the smooth Vlasov model, 
such as finite-$N$ effects, perturbing interactions, etc. 

To conclude, there is a single Vlasov equation, 
but there are various views of the distribution function whose evolution it is meant to describe. 
For a given plasma physics problem, which of these views, if any, should be considered~? 
This issue is usually overlooked, and the outcome of the vlasovian calculation is deemed relevant.

\subsection{Technical approach}
\label{sec:Status-technical}

The \emph{statistical} approach to deriving the Vlasov equation (see e.g.\ Appendix A in \cite{Ichi73}) 
starts from $N$-body dynamics, introduces the high-dimensional phase space $\Gamma = \RR^{6N}$,
and considers the Liouville equation for a statistical measure $\mathfrak{f}$ on $\Gamma$.\footnote{Given
   an initial data $z \in \Gamma$, the $N$-body equation of motion $\dot z = g(z)$ generates the evolution
   of the $6N$ degrees of freedom so that $z(t) = T_{t,0} (z_0)$ for initial data $z(0) = z_0$. 
   A statistical measure on $\Gamma$ is a distribution $\mathfrak{f}$,
   and the Liouville equation $\partial_t \mathfrak{f} + \partial_z \cdot (g \mathfrak{f}) = 0$ 
   transports an initial measure $\mathfrak{f}_0$ for $z_0$ 
   into an evolved measure $\mathfrak{f}_t = \mathfrak{f}_0 \circ T_{0,t}$ 
   (just by finding the preimage at time $0$ of the state at time $t$).
   For instance, for free motion, $z = (\mathbf{r}_1, \mathbf{v}_1, \ldots \mathbf{r}_N, \mathbf{v}_N)$,
   $g(z) = (\mathbf{v}_1, \mathbf{0}, \ldots \mathbf{v}_N, \mathbf{0})$ 
   so that $T_{t,0} z = (\mathbf{r}_1 + \mathbf{v}_1 t, \mathbf{v}_1, 
                                     \ldots \mathbf{r}_N + \mathbf{v}_N t, \mathbf{v}_N)$
   while $\mathfrak{f}(z, t) 
       = \mathfrak{f}(\mathbf{r}_1 - \mathbf{v}_1 t, \mathbf{v}_1, \ldots \mathbf{r}_N - \mathbf{v}_N t, \mathbf{v}_N, 0)$.
   }
This statistical measure may be interpreted as 
the probability distribution of, say, $\cN$ ``replicas'' ($\cN  \gg 1$) of the $N$-body system, 
and the symmetrized one-particle marginal\footnote{That is, 
   $f(\mathbf{r}, \mathbf{v}) 
   = N^{-1} \sum_{j=1}^N \int_\Gamma \delta ( \mathbf{r} - \mathbf{r}_j) \delta(\mathbf{v} - \mathbf{v}_j) \
                                                      \mathfrak{f}(\mathbf{r}_1, \mathbf{v}_1, \ldots \mathbf{r}_N, \mathbf{v}_N, t) 
                                                      \, \prod_{l = 1}^N \rmd^3 \mathbf{r}_l \rmd^3 \mathbf{v}_l$~:
   integration with the Dirac distribution generates the one-particle distribution for particle $j$, 
   and the sum symmetrizes over all particles.
   }
of $\mathfrak{f}$ obeys the first equation in the BBGKY hierarchy. 
One then identifies this symmetrized marginal with a one-particle measure
on the ``molecular'' $\mu$-space\footnote{This space would be the phase space of a particle,
   if the force field acting on it were known (though possibly time-dependent). 
   However, for the many-body problem, particles interact, 
   so that the phase space is $\Gamma = \otimes_{j=1}^N \RR^6$
   while $\mu$-space is just $\RR^6$ 
   (see notes 72 and 118 in \cite{Ehrenfest} for a historical gas theory analogue).
   }, 
and one expresses the force field generating the evolution by some integrals 
of the two-particle marginal of $\mathfrak{f}$. 
For the Vlasov--Poisson system, one then requires that the electric field in the Vlasov equation 
solves the Poisson equation where the source is 
the average of the individual particle distributions generated by all replicas.\footnote{The source 
   of $\mathbf{E}$ is not the $N^{-1} \sum_{j=1}^N \delta(\mathbf{r} - \mathbf{r}_j)$ given by a single
   plasma realisation, but the density $\int_{\RR^3} f(\mathbf{r},\mathbf{v}) \rmd^3 \mathbf{v}$
   obtained from an average over $\cN$ replicas.
   }
In this approach, the evolution of $f$ is thus subsidiary to the evolution of many replicas, 
each of which evolves independently of the $\cN-1$ other replicas, 
and the $f$ of interest in the Vlasov equation 
is then an average over $\cN$ replicas in the limit $\cN \to \infty$.
But how can ($\cN-1$) thought-experimental replicas drive the evolution of the single 
physically realized $N$-body system~? 
How should the field acting on a given particle in the physically observed plasma 
be the by-product of an average over $\cN$ \emph{Gedanken} plasmas~? 

The \emph{dynamical} approach to the Vlasov equation \cite{Spohn91,Kie14} starts with an 
actual system of $N$ bodies, interacting by instant-action-at-a-distance or via dynamical fields 
(see e.g.\  \cite{FiEl98} for a simple example). 
For undistinguishable particles, the data of $N$ points in $\mu$-space, 
namely a set\footnote{Not a \emph{sequence} where labels do matter !} 
  $$M = \{(\mathbf{r}_1, \mathbf{v}_1), \ldots, (\mathbf{r}_N, \mathbf{v}_N)\}$$ 
(thanks to Coulomb repulsion, 
no particles can be at the same position, hence this set counts exactly $N$ points), 
is equivalent to the counting measure 
$\rmd \mu^{(N)} 
  = \sum_{j=1}^N \delta (\mathbf{r}- \mathbf{r}_j) \delta (\mathbf{v} - \mathbf{v}_j) \,\rmd^3 \mathbf{r} 
                           \,\rmd^3 \mathbf{v}$ on $\RR^6$. 
When particles move with time, the measure also evolves, 
and for finite $N$ the evolution of $\mu^{(N)}$ provides all information on the motion of all particles,
as particles cannot swap their identity (exchanging two particle labels would require their trajectories to be discontinuous, 
or to meet at a same position with the same velocity -- which is ruled out by the dynamics). 
The measure $\mu^{(N)}$ determines the force field exactly as the $N$ particle data do, 
and this field generates the vector field according to which $\mu^{(N)}$ is transported. 
In order to handle the limit $N \to \infty$, 
it is convenient to consider the \emph{normalized} empirical measure
$N^{-1} \mu^{(N)}$, which is non-negative and verifies $N^{-1} \mu^{(N)}(\RR^6) = 1$. 
Indeed, this makes $\mu^{(N)}/N$ formally akin to a probability measure, 
and indeed one may interpret $N^{-1} \mu^{(N)}(\cA)$ as the fraction of the plasma particles 
which are in some subset $\cA \subset \RR^6$, 
or as the probability that a particle with randomly picked label\footnote{One shall not 
  confuse this ``label'' randomness, essentially related to particle undistinguishability
  in a single realization of the plasma,
  with the random choice of a replica out of $\cN$ possible realizations of the plasma.}
be in $\cA$.

The limit $N \to \infty$ makes sense formally for non-negative normalized measures, 
and indeed this space of measures can be equipped with various kinds of distances 
generating physically reasonable topologies \cite{Spohn91,MiMouWenn} 
to give an operational meaning to the notation  $\lim_{N \to \infty}$.
One such distance was considered by Kolmogorov and Smirnov to test 
the likelihood for sample random data to follow a given law, 
and it is used in convergence theorems for distributions \cite{Dudley}.

On the contrary, the limit $N \to \infty$ is ill-defined in the phase-space approach
involved in the Liouville equation, on which the BBGKY hierarchy relies. 
The very space in which the phase point (representing the $N$-particle system) evolves 
varies with $N$~: its dimension increases like $N$. 
Therefore, phase space $\Gamma$ is simply not the good stage for performing the limit. 

The derivation of the Vlasov equation in the (dynamical) measure approach is rather short 
(actually, shorter than the BBGKY-hierarchy based derivation) 
and both conceptually and physically clear,
provided the interaction is not too singular 
(with the statistical approach providing no better derivation in the singular case either). 
Key ingredients are 
\begin{itemize}
\item{the existence and uniqueness of solutions to individual particles' equations of motion 
   in a given, regular enough, force field, say $\mathbf{E}(\mathbf{r},t)$ 
   (with e.g.\ Lipschitz regularity,
   i.e.\ $\norm{\mathbf{E}(\mathbf{r},t) - \mathbf{E}(\mathbf{r}',t)} \leq K(t) \norm{\mathbf{r} - \mathbf{r}'}$),
   }
\item{by duality\footnote{This is just 
      using the fact that $\mu(T_{t,0} (y) \in \cA) = \mu(y \in T_{0,t} \, \cA)$, because 
      the space of measures $\mu$ is dual to the space of observables $q$,
      in the linear algebraic interpretation of the formula $\int_{\RR^6} q(y) \, \rmd \mu(y)$
      \cite{Appel}.
      }
   and because the single-particle dynamics in the given field 
   is measure-preserving in the ``molecular'' $\mu$-space, 
   the existence and uniqueness of evolution of measures 
   in the same given, regular enough, force field $\mathbf{E}(\mathbf{r},t)$,
   }
\item{regularity of the force field generated by arbitrary measures 
   (which fails for Coulomb field with point sources),
   }
\item{a self-consistency argument to formulate the Vlasov equation coupled with the fields 
   as a fixed point problem in a suitable (Banach) function space 
   -- much like the standard technique used for solving iteratively ordinary differential equations. 
   }
\end{itemize}

The measure approach reaches further than just proving 
that the limit $N \to \infty$ commutes with time evolution. 
It actually shows that, 
given initial measures $\mu_1(0)$ and $\mu_2(0)$ on $(\mathbf{r},\mathbf{v})$-space,
which need not be empirical nor absolutely continuous,
the Vlasov equation defines unique time-dependent measures $\mu_1(t)$ and $\mu_2(t)$
and that the distance between the evolved measures is controlled by their initial distance
and a constant $C$ depending on $\mu_1(0)$ and $\mu_2(0)$,
\begin{equation}
  \mathrm{dist}(\mu_1(t), \mu_2(t)) 
  \leq 
  C' \, \mathrm{dist}(\mu_1(0), \mu_2(0)) \, \rme^{C \abs{t}}  .
  \label{distmu}
\end{equation}
Here $C'$ may occur for technical reasons, and
$C$ is an upper estimate for the largest Lyapunov exponent,
controlling the rate at which particles separate in the fields $\mathbf{E}_1$ and $\mathbf{E}_2$ 
generated by the matter distributions described by $\mu_1$ and $\mu_2$
(see \cite{FiEl98,EEbook} for a smooth interaction example). 
This estimate implies that, to ensure
$\mathrm{dist}(\mu^{(N)}(t)/N, \mu_f(t)) \sim \delta$ 
with $\mathrm{dist}(\mu^{(N)}(0)/N, \mu_f(0)) \sim N^{-c}$ (with $c = 1/2$ or $1$, say),
one needs an initial accuracy corresponding to some $N \sim (C' / \delta)^{1/c} \exp(C \abs{t} /c)$,
which is too demanding when the times of interest 
exceed a few Lyapunov e-folding times. 

As physicists are more interested in estimating specific observables, 
say local current densities or electric field energy density, 
estimates like (\ref{distmu}) provide upper bounds 
on the errors due to approximating $\mu_1(0)$ with $\mu_2(0)$. 
However, actual discordances between the values of specific observables 
are generally much smaller -- for instance, if $\mu_1(0)$ and $\mu_2(0)$ describe 
two stationary solutions, the observables are constants of the motion.

\section{Vlasov--Poisson evolution}
\label{sec:VP}

While the dynamical foundation of the Vlasov equation, in the previous section, stressed the motion of particles,
we must stress that some important results on the Vlasov--Poisson system make no use of particle trajectories. 
In this section, we first indicate how the smooth solutions to the Vlasov--Poisson system 
may involve non-trajectorial concepts, 
and then comment on some aspects of the system which may, or need to, involve trajectories.

\subsection{Without trajectories}
\label{sec:VlaPoi-notraj}

As a partial differential equations system, the Vlasov--Poisson model is well-posed and much studied
(see e.g.\  \cite{Rendall}). Theorems on solutions existence rely on some regularity in initial data,
typically  $f$ has compact support and bounded, continuous 
derivatives $\partial_\mathbf{r} f$, $\partial_\mathbf{v} f$. 
Proofs involve estimates for the spatial density 
$\rho (\mathbf{r}, t) = \int_{\RR^3} f(\mathbf{r},\mathbf{v},t) \, \rmd^3 \mathbf{v}$ 
in various $L^p$ norms, along with estimates for the electric field $\mathbf{E}$ 
(whose $L^2$ norm is proportional to the total potential energy) and its gradients,
which involve norms of $\partial_\mathbf{r} \rho$. 
Moments of $f$ (and of course of $\rho$), as well as the upper bound on particle velocity
$P(t) = \sup \{ \abs{\mathbf{v}}~: f(\mathbf{r},\mathbf{v},t) > 0 \textrm{ for some } \mathbf{r} \}$, 
play a role in the analysis of the Vlasov--Poisson system. 

It may occur that gradients of $f$ get steep, due to filamentation in $\mu$-space. 
Yet, gradients of $\rho$ or of $\bar f(\mathbf{v},t)~:= \int_{\RR^3} f(\mathbf{r},\mathbf{v},t) \, \rmd^3 \mathbf{r}$ 
may be better controlled, and even decay in some sense to zero,
as e.g.\ in the nonlinear theory of Landau damping \cite{MV09,MV10,Villani14,BedMasMou13}. 

The analysis of continuous solutions to the Vlasov--Poisson system 
does not rely only on the more usual physical quantities like 
$\rho$ or $\mathbf{j}(\mathbf{r},t) = \int_{\RR^3} \mathbf{v} f(\mathbf{r},t) \, \rmd^3 \mathbf{v}$, 
and on standard global invariants like total momentum or energy. 
It also takes advantage of Casimir invariants, which may be expressed in terms of the functions 
\begin{equation}
  C_a [f] (t)~:= 
  \int_{\RR^6} 1(f(\mathbf{r},\mathbf{v},t) > a) \, \rmd^3 \mathbf{r} \, \rmd^3 \mathbf{v}
\end{equation}
where $1(A) = 1$ if $A$ is true and $1(A) = 0$ otherwise. 
Note that $C_a$ is a decreasing function of $a$, with $C_a = + \infty$ for any $a < 0$,
and $C_0$ being the measure of the support of $f$.
These functions $C_a$ measure the level sets of $f$ 
and provide a tool for calculating other functionals like
$\norm{f}_{L^1} = \int_0^{\infty} C_a[f] \, \rmd a$, 
$\norm{f}_{L^p}^p = p \int_0^{\infty} a^{p-1} \,ÊC_a[f] \, \rmd a$, or 
$\int_{\RR^6} f(\mathbf{r},\mathbf{v}) \ln (f(\mathbf{r},\mathbf{v})/c)  \rmd^3 \mathbf{r}  \rmd^3 \mathbf{v} 
     = \int_0^{\infty} (1 + \ln (a/c)) C_a[f]  \rmd a$. 

The Vlasov equation preserves all Casimir invariants, 
and a physical interpretation for these conservation laws 
is that vlasovian evolution must be invariant under the group of particle relabelings. 
Such a group is discrete for the $N$-body system (and then $C_a$ is undefined), 
but it is continous for smooth $f$, 
and one expects it to generate integral invariants following Noether's theorem. 
Adding to the total energy a suitably chosen Casimir invariant enabled 
proving the stability of some spherically symmetric equilibria 
of the gravitational Vlasov--Poisson system \cite{Mouhot11}. 

In numerical simulations and in simple analytic models, 
the case where $C_a$ is a simple step function occurs for waterbag distributions \cite{BertrandFeix68},
equal to $h$ on a domain $\cA(t)$ with measure $1/h$, and vanishing outside $\cA(t)$
(then $C_a = 0$ for $a \geq h$ and $C_a = 1/h$ for $0 \leq a < h$).

\subsection{With trajectories}
\label{sec:VlaPoi-traj}

Particle trajectories are instrumental in understanding the Vlasov equation 
because this equation transports the distribution $f$ along the characteristics of the Vlasov operator. 
Good control on particle trajectories is crucial e.g.\ 
to proofs of the existence of solutions globally in time \cite{Pfaff92,Horst,Schaeffer91}
(using the Lipschitz regularity of the electric field generated by the distribution function) 
and to the proof of Landau damping in the nonlinear regime \cite{MV09,MV10,Villani14,BedMasMou13}. 

Estimates on individual particle trajectories in small field limits are obtained 
for the fields generated by the Vlasov--Poisson dynamics, and provide further insight in the latter,
using estimates for velocity moments of the distribution function \cite{LionsPerthame,Pallard11}.  
Estimates on particle trajectory crossings are also used to construct Lyapunov functionals 
for the evolution from ``small'' initial data in order to assess the decay to spatial uniformity 
and zero electric field in long times, 
and the ``asymptotic completeness''\footnote{This issue is
   whether the nonlinear Vlasov flow can or cannot be approximated by the free streaming flow 
   in the limit $t \to \infty$.} 
of dynamics \cite{ChaeHa06,ChoiHa11}.

Incidentally, Mouhot and Villani also stress the stabilizing role of plasma echoes 
(a particle effect, associated with bunching)  
for the nonlinear theory of Landau damping. 
However, characteristics (i.e.\ trajectories) are not considered individually~; 
rather, smoothness of $f$ is important, and nonlinear particle motion does not break it for finite time -- 
while mixing in $(\mathbf{r},\mathbf{v})$ space generates small scale oscillations 
(ripples) which are homogenized in the long run. 

On the other hand, particle trajectories may be ignored when considering stationary solutions, 
see e.g.\ the orbital stability problem solved by Lemou, M\'ehats and Rapha\"el (see \cite{Mouhot11} for a review).
Nor do they appear explicitly in the discussion of Bernstein-Greene-Kruskal modes and trapping 
in space-periodic Vlasov--Poisson system \cite{LinZeng}.

Note that all these statements refer to the Vlasov--Poisson system 
as a partial-differential deterministic evolution equation system,
in which any particle is following a trajectory defined by the Vlasov equation characteristics. 
It is not analysed with probabilistic methods.

\section{Direct many-body dynamics}
\label{sec:Nbody}

A finite-$N$ approach always stresses trajectories. 
Analogues of the stability theorems for Landau damping in the nonlinear regime would be a 
Kolmogorov--Arnol'd--Moser theorem \cite{BedMasMou13} for global-in-time stability, 
or a Nekhoroshev-type theorem for stability over exponentially long times 
(say $\exp(a N^b)$ for some $a, b > 0$). 
This is indeed a challenging target for mathematical analysis (see the recent tutorial \cite{Villani14}).

Interestingly, (partial-differential-equations-based) Landau damping theorems 
involve periodic boundary conditions, 
so that the electrons may be viewed as immersed in a neutralizing background (the ``jellium'' model)
-- while results on asymptotic completeness involve small initial data in an infinite space. 

For periodic jellium, with finitely many bodies, interacting through repulsive Coulomb force,
no two particles can be simultaneously at the same point. 
The equations of motion are well-posed,
and solutions to the initial value problem (Cauchy problem) exist for all times.\footnote{For actual
   plasmas, with ions and electrons, Coulomb attraction permits that particles come 
   arbitrarily close, with potential energy diverging to $-\infty$ while kinetic energy diverges to $+\infty$.
   This makes their $N$-body description even more difficult. 
   The similar difficulty in celestial mechanics is well known \cite{Falcolini}.
   }  
The continuum limit, where $N \to \infty$ along with $\mu^{(N)}/N$ approaching 
an absolutely continuous measure with a smooth density $f$, 
is however difficult because the limit imposes that some particles come arbitrarily close
to each other, and the characteristic time scales for their relative motion become arbitrarily short. 
The short-range singularity of the Coulomb potential is too strong for the usual rigorous derivations of the 
Vlasov equation to apply.   
Kiessling \cite{Kie14} exposes a new approach along with a review of the state of the art.

Without relying on the Vlasov description, one can use 
the $N$-body description directly to understand some fundamental physical effect in plasmas.
This issue was recently revisited \cite{EDE13} for $N$ electrons (with mass $m_\rme$ and charge $-e$)
in a cube with side $L$ with periodic boundary conditions (the latter play the role of a neutralizing background).
For large $N$, when the particle positions are not too far from a spatially uniform distribution, 
and velocities are close to a distribution $f_0(\mathbf{v})$,
one separates the motion of a particle into a ballistic 
approximation,  $\mathbf{r}_l^{(0)}(t) = \mathbf{r}_{l0} + \mathbf{v}_{l0} t$,
and a correction $\delta \mathbf{r}_j(t) = \mathbf{r}_j(t) - \mathbf{r}_l^{(0)}(t)$ due to the interactions. 
Perturbative analysis\footnote{See 
   appendix B in \cite{EDE13} for the outline of a nonlinear version.}  
shows that the dominant contribution to their interaction 
is not given by the full Coulomb potential 
(with Fourier components $ \tilde{\varphi}(\mathbf{m},t)
  = - \sum_{j = 1}^N e / (\epsilon_0 \norm{\mathbf{k}_\mathbf{m}}^2) 
     \exp(- \rmi \mathbf{k}_{\mathbf{m}} \cdot \mathbf{r}_j(t))$
for $\mathbf{k}_\mathbf{m} =  2\pi\mathbf{m}/L$ 
with $\mathbf{m} \in \ZZ^3 \setminus \{\mathbf{0}\}$),
but only the Debye-shielded potential 
$\delta \Phi(\mathbf{r},t) 
= \sum_j  \delta \Phi(\mathbf{r} - \mathbf{r}_j(0) - \dot{\mathbf{r}}_j(0) t,\dot{\mathbf{r}}_j(0))$,
where
\begin{equation}
  \delta \Phi (\mathbf{r},\mathbf{v})
  = - \frac{e}{L^3 \epsilon_0} \sum_{{\mathbf{m}} \neq {\mathbf{0}}}
      \frac{\exp(\rmi \mathbf{k}_{\mathbf{m}} \cdot \mathbf{r})}
           { \norm{\mathbf{k}_{\mathbf{m}}}^2 \, 
             \epsilon(\mathbf{m},\mathbf{k}_{\mathbf{m}} \cdot \mathbf{v} + \rmi \varepsilon) }
\label{phi}   
\end{equation}
and $\epsilon$ approaches\footnote{The $\rmi \varepsilon$ prescription 
   stems from inverting the Laplace transform as
   the integral in $\epsilon$ is undefined for real-valued
   $\omega = \mathbf{k}_{\mathbf{m}} \cdot \mathbf{v}$.
   As this diverging $\epsilon$ occurs in the denominator in (\ref{phi}), 
   however, the finite $N$ version does not need the $\rmi \varepsilon$.
   A similar discussion occurs in the analysis of discrete analogues of 
   van Kampen-Case modes \cite{EEbook}.}
the usual dielectric function 
\begin{equation}
  \epsilon(\mathbf{m},\omega)
  = 1 -  e^2 / (L^3 m_\rme \epsilon_0)
     \int  (\omega -\mathbf{k}_{\mathbf{m}}  \cdot \mathbf{v})^{-2} f_0(\mathbf{v}) 
          \,  \rmd^3 \mathbf{v}
\label{eps}
\end{equation}
in the $N \to \infty$ limit.

While Debye shielding of probes or test charges is well known in the kinetic theory of plasmas, 
this result brings forward the direct appearance of Debye shielded Coulomb potential 
in the dynamics -- which is not usually stressed in classical Vlasov--Poisson theory. 
Moreover, the faster decay of the shielded force enables one to compute transport phenomena
in terms of particle ``binary collisions'' over the whole range of impact parameters \cite{EED14},
while the classical collisional models were constructed separately 
for large and small impact parameters. 
In the same calculations, the collective response of particles to any one of them 
also ``dresses'' the latter, so that the effective description of the plasma deviates 
simultaneously from both the Vlasov equation (stressing now $N$ bodies) 
and from the Poisson equation (through shielding and dressing). 

The potential (\ref{phi}) lends itself easily to Laplace analysis in time \cite{EDE13}. 
In the $N \to \infty$ limit, it then also approaches formally (outside its sources) 
the potential given by the standard expression involving the Landau contour calculation,
for both the unstable (growing wave) case and the damping case. 
The twist in this derivation of Landau damping in the linear regime 
is that it involves not the full Coulomb interaction 
but its (perturbatively constructed) shielded version. 
This comes as a surprise, for screening generates a ``short range'' effective model~;
yet the shielded potential is still Coulomb-singular at its source,   
so that the standard derivations of the Vlasov equation 
do not apply better than for the full Coulomb interaction.

\section{Final remarks}
\label{sec:Final}

Let us first recall that finite-$N$ dynamics may always depart from smooth Vlasov solutions 
on the characteristic time scales of instabilities, as noted in eq.~(\ref{distmu}).
Apart form this obvious example, it is known that
for interactions smoother than Coulomb's, finite-$N$ distributions can indeed 
remain quasi-stationary for extremely long times (see \cite{Levin14} for a recent review), 
even if the finite-$N$ model does not admit corresponding periodic solutions
\cite{El01}.
Moreover, the large, finite $N$ regime of near-equilibrium distributions, 
as occurs in the physical problem motivating research on orbital stability, 
is of interest for the Newton gravitation interaction \cite{Mouhot11},
where it also branches to the elegant special solutions to the $N$-body dynamics 
with symmetries known as ``choreographies'' \cite{Montgomery10}. 
Similar results in the Coulomb case relate to the motion of cold ion clouds 
or non-neutral plasmas \cite{Kie09,Dubin99}.

As physicists have since long understood that finite-$N$ systems are never completely described
by classical Vlasov--Poisson theory, long-time evolutions of plasmas are often described 
using other kinetic models in the $N \to \infty$ limit. 
Rigorous derivations of such models, 
like the Landau equation or the Balescu--Lenard--Guernsey equation,
are a challenge for this century (see  \cite{Lan10} and the truly short ch.~6 in \cite{Spohn91}). 
Actually, corrections to the Vlasov model often go by the name of ``collisions'',
and we noted in Sect.~\ref{sec:Nbody} 
how surprising they may be even in their simpler instances \cite{EDE13,EED14}.
Collisions do not necessarily account for the graininess effects due to finite $N$ \cite{FirpoPoleni}.

\medskip

It is a pleasure for YE to acknowledge stimulating discussions with G.~Belmont, F.~Califano, 
C.~Krafft, G.~Manfredi, Ph.~Morrison, F.~Pegoraro, F.~Valentini and participants to Vlasovia 2013 in Nancy. 
The authors are grateful to L.~Cou\"edel for a critical reading of the manuscript, and to 
their colleagues in Marseilles for fruitful comments. 

%
%

\end{document}